\documentclass[%
 reprint,
nofootinbib,
nobibnotes,
bibnotes,
 amsmath,amssymb,
 aps,
prb,
]
{revtex4-2}

\usepackage{graphicx}
\usepackage{dcolumn}
\usepackage{bm}
\usepackage{natbib}
\usepackage{color}

\begin{document}

\preprint{APS/123-QED}

\title{
Ferromagnetic Crossover within the Ferromagnetic Order of U$_{7}$Te$_{12}$}

\author{Petr Opletal}
\author{Hironori Sakai}
\author{Yoshinori Haga}
\author{Yoshifumi Tokiwa}
\author{Etsuji Yamamoto}
\author{Shinsaku Kambe}
\author{Yo Tokunaga}
%
\affiliation{%
Advanced Science Research Center, Japan Atomic Energy Agency, Tokai, Ibaraki 319-1195, Japan
}%





\begin{abstract}
We investigate the physical properties of a single crystal of uranium telluride U$_{7}$Te$_{12}$.
We have confirmed that U$_{7}$Te$_{12}$ crystallizes in the hexagonal structure with three nonequivalent crystallographic uranium sites.
The paramagnetic moments are estimated to be approximately 1 $\mu_{\rm B}$ per the uranium site, assuming a uniform moment on all the sites.
A ferromagnetic phase transition occurs at $T_{\rm C}=48$ K, where the in-plane magnetization increases sharply, whereas the out-of-plane component does not increase significantly.
With decreasing temperature further below $T_{\rm C}$ under field-cooling conditions, the out-of-plane component increases rapidly around $T^{\star}=26$ K.
In contrast, the in-plane component hardly changes at $T^{\star}$.
Specific heat measurement indicates no $\lambda$-type anomaly around $T^{\star}$, so this is a cross-over suggesting a reorientation of the ordering moments or successive magnetic ordering on the part of the multiple uranium sites.
\end{abstract}

\maketitle

\section{Introduction}
Uranium 5$f$ electrons play an important role in bringing out various solid-state properties.
In comparison to 4$f$ electron states in lanthanide compounds, which are strongly localized, 5$f$ electron states extend into space \citep{Sechovsky_1998}.
This leads to inter-site interaction with direct overlap by 5$f$ electron states of the nearest uranium atoms, and/or hybridization with conduction electron states of ligand atoms.
Meanwhile, the intrasite coupling with 6$d$ electron states is also important for microscopic understanding.
To investigate the role of 5$f$ electrons in physical properties, it is important to systematically study a group of compounds constructed from the same elements.
In this sense, uranium chalcogenides are promising research targets, which
exhibit various physical properties, such as, piezomagnetism in UO$_2$ \citep{Antonio2021}, metal-insulator crossover induced by magnetic field in $\beta$-US$_2$ \citep{IkedaS:JPSJ78:2009, Sugiyama2011}, and spin-triplet superconductivity recently discovered in UTe$_{2}$ \citep{Ran2019a}.

Especially, there has been a growing interest in the adjacent phase of UTe$_2$, which is not a line phase, including nonstoichiometry caused by U deficiency \citep{Haga2022}.
In addition, U$_7$Te$_{12}$ can be grown with stoichiometric UTe$_{2}$ by the molten salt flux method \citep{PhysRevMaterials.6.073401}.
U$_{7}$Te$_{12}$ is reported to crystallize in non-centrosymmetric hexagonal Cr$_{12}$P$_{7}$-like structure (space group $P\overline{6}$, No. 174) \citep{Breeze1971, Tougait1998}.
This structure is characterized by three nonequivalent uranium sites -- U1 with multiplicity 1 and U2 and U3 with multiplicity 3.
Only reported data for U$_{7}$Te$_{12}$ were obtained on polycrystalline samples showing ferromagnetic transition at 52 K \citep{Suski1972, Tougait1998}.
Although anisotropic magnetism is expected from the crystal structure, the details of this magnetism are unknown.
Moreover, it was reported as a semimetallic compound \cite{Tougait1998}. However, no data on temperature-dependent resistivity has been presented until now.

In this study, we prepared single crystals of U$_{7}$Te$_{12}$ by chemical vapor transport (CVT) and molten salt flux methods.
Single crystals enable us to examine the anisotropic physical properties of this compound.
Two different magnetic anomalies are observed through magnetic, electrical transport, and heat capacity measurements: one is a ferromagnetic transition at $T_{\rm C}=48$ K, and another is a ferromagnetic crossover at $T^{\star}=26$ K.
The former ferromagnetic ordering is characterized by an increase of the $a$-axis magnetization, while the latter crossover is characterized by an increase of $c$-axis magnetization.
We have also revealed that the material exhibits semimetallic conductivity, of which resistivity slightly decreases below $T_{\rm C}$ and conversely increases below $T^{\star}$.
It may be a half-gapped semimetallic state induced by the ferromagnetic crossover.
We discuss the possible nature of these magnetic anomalies and their connection to nonequal uranium sites.

\section{\label{sec:level1}Experimental methods}
Single crystals of U$_{7}$Te$_{12}$ were prepared by the CVT method  with NH$_{4}$Cl as a transport agent.
Precursor elements of etched U metal and Te (Te 6N purity produced by Rare Metallic Co. Ltd.) in a 1:1.5 ratio were used and the quartz tube for growth was etched and smoothed inside by hydrofluoric acid.
Precursor elements were sealed inside the quartz tube under high vacuum (10$^{-5}$ mbar) with approx. 0.04 mg/cm$^{3}$ of NH$_{4}$Cl.
During the evacuation, the whole quartz tube was heated up to 100$^{\circ}$C to desorb any water in NH$_{4}$Cl.
The sealed tube was placed in a two-zone horizontal electric tube furnace (ARF2-370-50KC, Asahi-rika Co., Ltd.).
The entire tube was heated up to 950$^{\circ}$C  for 24 hours to allow the reaction of contents.
An opposite gradient was applied to clean the growth zone for three days.
Following that, the gradient was flipped with the lower temperature - 850$^{\circ}$C  at the growth zone (etched by hydrofluoric acid) and higher temperature - 950$^{\circ}$C at the charge zone and was left for 10 days.
We were also able to prepare a single crystal by the molten-salt method as described in \cite{PhysRevMaterials.6.073401}.

Single crystals grown by the molten-salt method tend to be prism-like with a hexagonal base growing along the c-axis with a thickness of 0.5 to 1 mm and a length of 2 to 6 mm.
Single crystals obtained by the CVT method tend to have a block shape with dimensions ranging from 2 to 4 mm.
Samples obtained by both methods have dark metallic color.

The crystal structure was determined by single crystal diffraction at room temperature using graphite monochromated Mo $K_{\alpha}$ radiation.
The scattered X-ray beam was recorded on an image plate detector (R-AXIS RAPID, Rigaku Corp.).
Absorption correction with the empirical method was applied prior to the structural solution. Structural solution and crystallographic parameters fitting were performed by the SHELX program.
The absence of any different phases was checked by electron-probe microanalysis using wavelength-dispersive spectrometers installed in a scanning electron microscope (JXA-8900, JEOL Ltd.).

To measure the physical properties, samples were cut from larger single crystals by an electric spark saw and polished.
Magnetic properties were measured by a commercial superconducting quantum interference device magnetometer (MPMS, Quantum Design) in the temperature range of 4.2 K to 300 K and a maximum magnetic field of 7 T.
The obtained data were not corrected for the demagnetization effect.
We estimate the demagnetization field to be below 0.05 T for the highest magnetic moment. 
The heat capacity was measured by relaxation method by two $\tau$ models in Physical Properties Measurement System (PPMS, Quantum Design) from 1.8 K up to 300 K.
Electrical resistivity was measured by the four-probe method in PPMS.       

\section{\label{sec:level2}Results}

\subsection{Crystal structure}

\begin{figure}[!tbh]
\includegraphics[width=8.5cm]{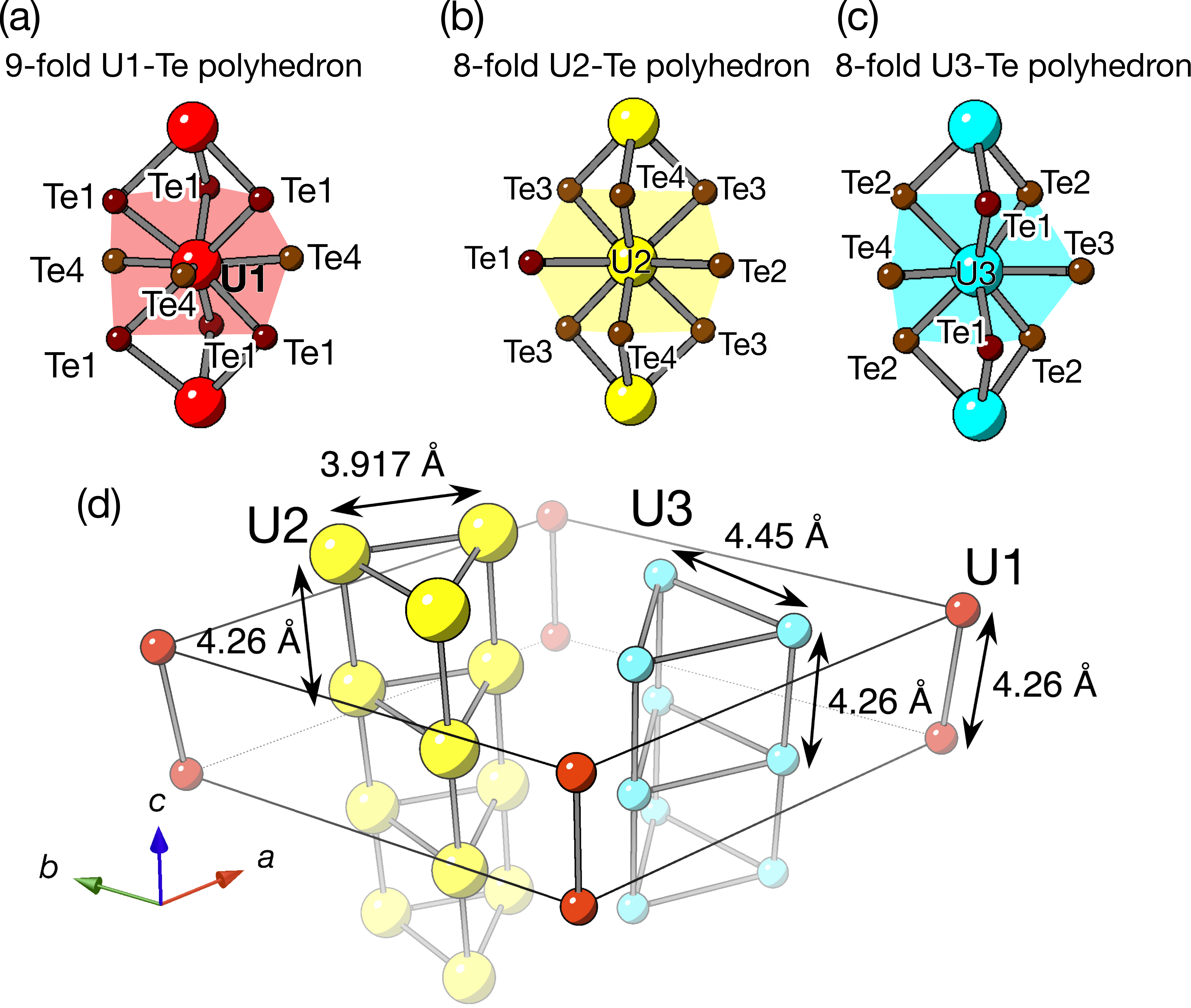}
\caption{\label{fig:structure} Crystal structure of U$_{7}$Te$_{12}$ with coordination polyhedra for (a) U1 in tricapped trigonal prismatic and (b) U2 and (c) U3 in bicapped trigonal prismatic geometry. (d) Crystal structure only with uranium  (for better clarity) highlighting triangular prism of U2 and U3. Distances between uranium atoms are shown.}
\end{figure}

\begin{table*}[]
\caption{\label{tab:table2}
 Wyckoff notation, local symmetry, atomic coordinates, equivalent isotropic displacement parameter, and site occupancy for each atomic position in U$_{7}$Te$_{12}$ which are obtained from the single crystal X-ray refinement.
}
\begin{ruledtabular}
\begin{tabular}{cccccccc}
\textrm{atom}&
\textrm{site}&
\textrm{local symm.} &
$x$ & $y$ & $z$ &
$B_{\mathrm eq}$ (\AA$^2$)  & occupancy\\
\colrule
U1 & $1a$ & $\overline{6}..$ & 0 & 0 & 0 & 0.53(3) & 1 \\
U2 & $3k$ & $m..$ & 0.17296(11)  & 0.46524(10)  & 1/2 & 0.58(3) & 0.983(6)  \\
U3 & $3j$ & $m..$ & 0.43295(10)  & 0.26651(11)  & 0 & 0.56(2) & 1 \\
Te1 & $3k$ & $m..$ & 0.21589(19)  & 0.2081(2)  & 1/2 & 0.57(4) & 1 \\
Te2 & $3k$ & $m..$ & 0.52583(19)  & 0.1345(2)  & 1/2 & 0.59(3) & 1 \\
Te3 & $3j$ & $m..$ & 0.3765(2)  & 0.4949(2)  & 0 & 0.71(4) & 1 \\
Te4 & $3j$ & $m..$ & 0.0173(2)  & 0.2649(2)  & 0 & 0.72(5) & 0.955(10) \\
\end{tabular}
\end{ruledtabular}
\end{table*}

We have confirmed U$_{7}$Te$_{12}$ crystallizes in Cr$_{12}$P$_{7}$-like structure (space group $P\overline{6}$, No. 174) with crystal lattice parameters $a = 12.277(3)$ \AA\ and $c = 4.2646(5)$ \AA.
The resulting coordinates of the atoms are presented in Table \ref{tab:table2}.
The structure was determined with final agreement factor $R_1$ = 3.05\%.
If the atomic occupancies on the U2 and Te4 sites were fixed as 1, the obtained equivalent displacement coefficients ($B_{\rm eq}$) were too large.
Alternatively, if these occupancies were treated as fitting parameters, the occupancies were slightly reduced. However the $B_{\rm eq}$ values become reasonable, and then the $R_1$-value became slightly reduced.
The resulting displacement factor of U2 sites remains large and qualitatively consistent with the result by Tougait {\it  et al.} \cite{Tougait1998}
As shown in Fig. 1(b), U2 is located in the polyhedron formed by Te3 and Te4, where U2 is much closer to Te4 leaving more space in the opposite direction.
The atomic displacement would likely be larger in this direction, resulting in a large displacement factor.

As shown in Fig. \ref{fig:structure}, two different coordination polyhedra of uranium and tellurium are seen in U$_{7}$Te$_{12}$.
The coordination polyhedra for U1 sites is tricapped trigonal prismatic (Fig. \ref{fig:structure}(c)), while those for U2 and U3 sites are bicapped trigonal prismatic geometry (Figs. \ref{fig:structure}(d) and \ref{fig:structure}(e)).
Therefore, two types (at least) of crystalline electric field (CEF) effects are expected to exist for different polyhedral U sites.

At first glance, the polyhedral arrays of U2 and U3 appear similar, but their polyhedral sequences are quite different as follows.
In the U$_7$Te$_{12}$ structure, the shortest distance between uranium sites is the U2-U2 distance of 3.917 \AA\ which is longer than Hill's limit \citep{Hill1970}.
This shortest U2-U2 corresponds to a side of the U2 triangle in the basal plane (Fig. \ref{fig:structure}).
Since each U2 site is aligned in a straight line in the $c$-axis direction, every $c = 4.2646$ \AA\ of the lattice constant, a geometrical frustration effect may exist in the U2 prism if antiferromagnetic interactions were dominant.
Meanwhile, the shortest distance in the U3 prisms corresponds to that along the $c$-axis, i.e., it is equal to the lattice parameter of $c$, whereas the side in the U3 triangle is equal to 4.45 \AA. The shortest distance in U1 prism is the same as in U3, but for U1 no triangle exists.
In between different uranium sites, the shortest one is found between U1-U3 equal to 4.67 $\mathrm{\AA}$.
The distances between U1-U2 and between U2-U3 are more than 5 $\mathrm{\AA}$, as a result of the stacking sequence of U2 being shifted by $c/2$ as shown in Fig. \ref{fig:structure}(a).

\subsection{Magnetic properties}
\begin{figure}[!tbh]
\includegraphics[width=8.5cm]{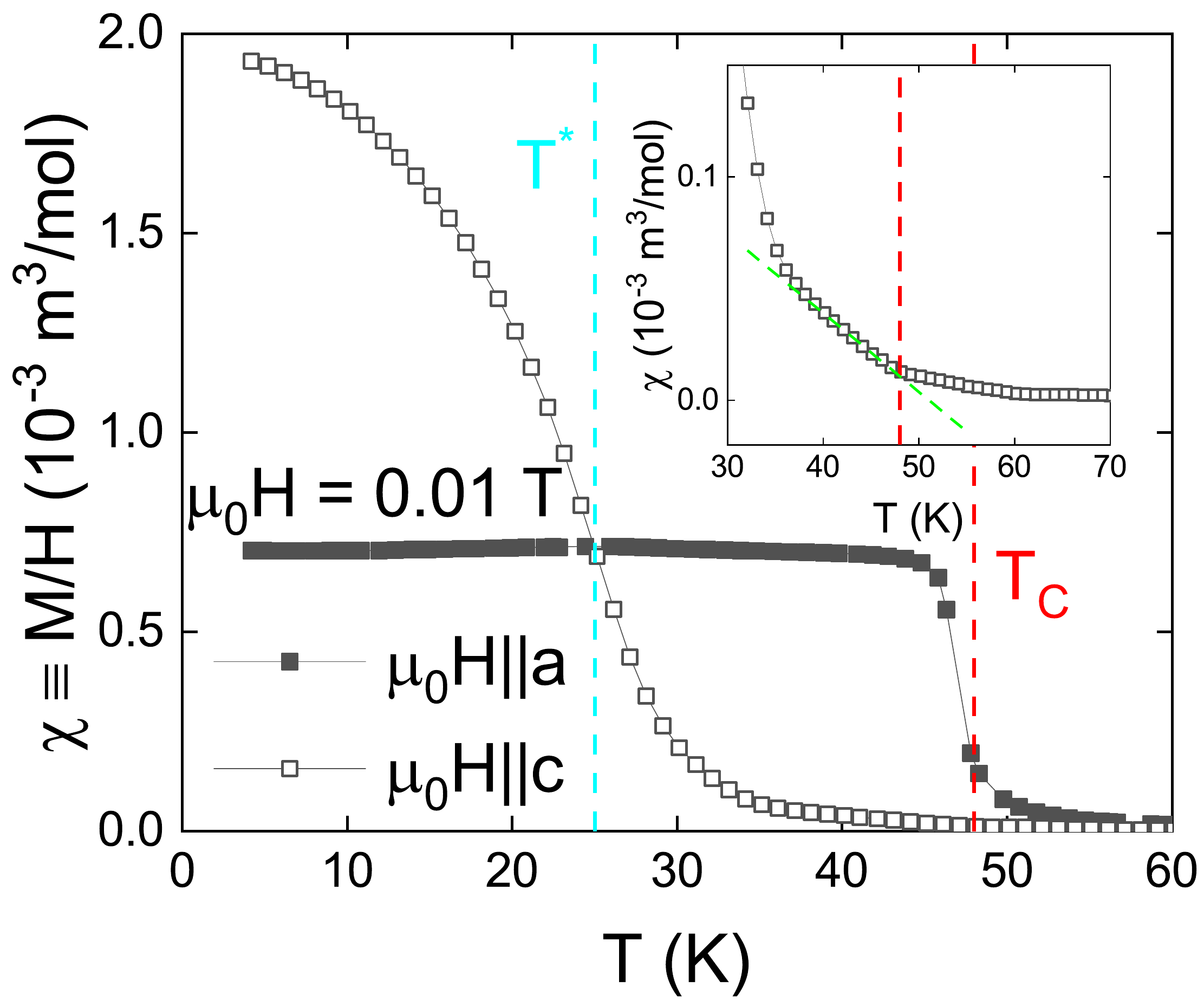}
\caption{\label{fig:M_T_a}
Temperature dependence of magnetic susceptibility defined as $M/H$ for U$_{7}$Te$_{12}$ with an application of $\mu_0H=0.01$ T along the $a$-axis and $c$-axis.
The data were taken under field-cooled (FC) conditions.
Closed and open squares show data obtained from the field applied along the $a$-axis and $c$-axis, respectively.
The inset shows detailed behavior close to ferromagnetic transition at 48 K for the field applied along the $c$-axis.
}
\end{figure}

\begin{figure}[!tbh]

\includegraphics[width=9cm]{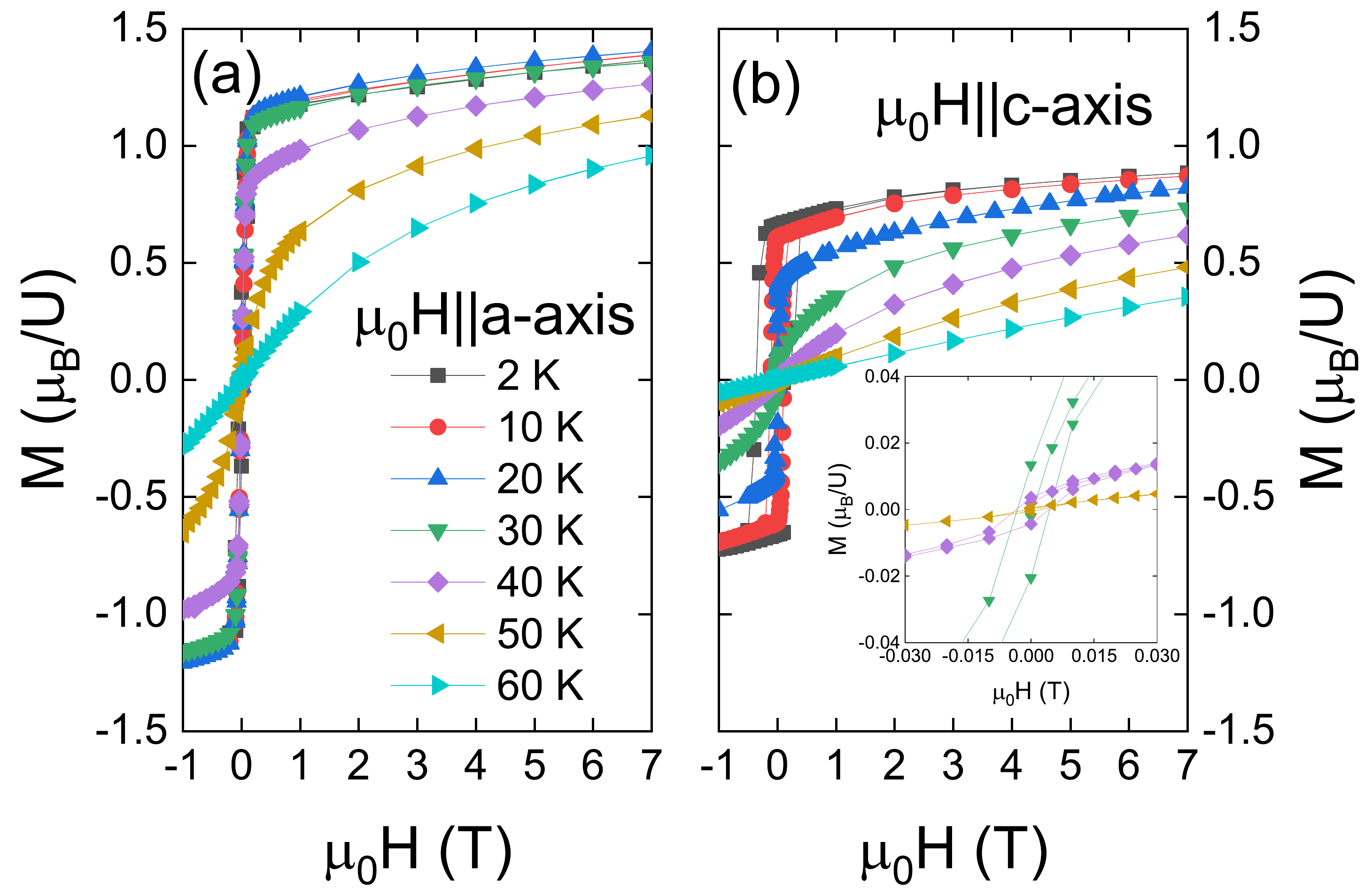}
\caption{\label{fig:M_H}
Left figure shows magnetic curves at different temperatures for the field applied along the $a$-axis. Right figure shows magnetic curves at different temperatures for the field applied along the $c$-axis with an inset showing detail of magnetic curves around the zero field region.}
\end{figure}

\begin{figure}[!tbh]
\includegraphics[width=8.5cm]{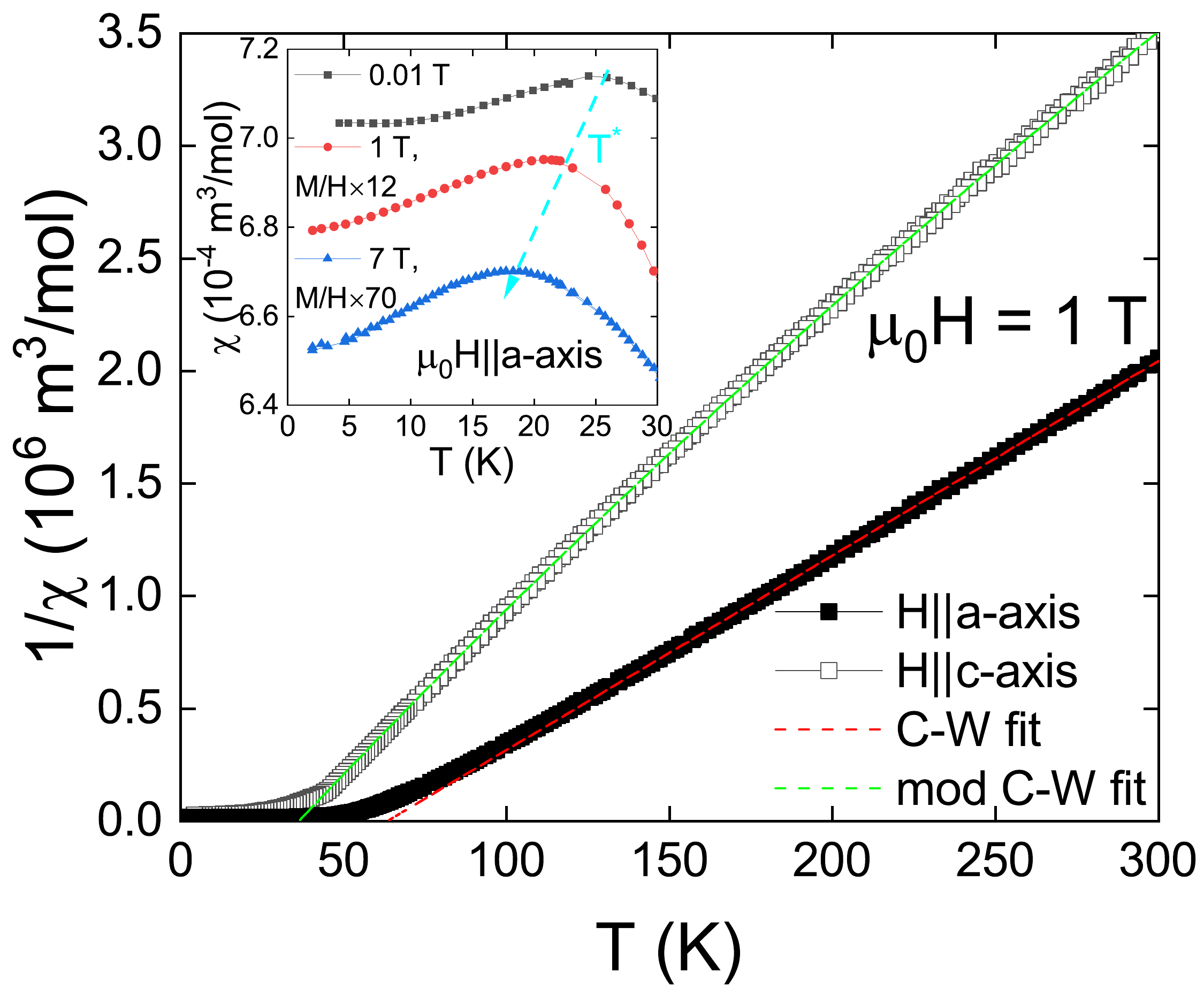}
\caption{\label{fig:susc} Temperature dependence of inverse magnetic susceptibility for magnetic field applied along the $a$-axis and $c$-axis. The fit to data by Curie-Weiss law and modified Curie-Weiss law are shown as dashed lines (for more information see text). The inset shows the low-temperature data of $M/H$ for different magnetic fields applied along the $a$-axis. Data at 1 T and 7 T are scaled and shifted for clarity. 
}
\end{figure}

Figure \ref{fig:M_T_a} shows the temperature dependence of magnetic susceptibility $\chi$ which is defined as the magnetization ($M$) divided by the applied magnetic field ($H$) along the $a$- and $c$-axes for U$_7$Te$_{12}$.
The data shown in Fig. \ref{fig:M_T_a} were taken in the field cooled (FC) condition.
As seen in Fig. \ref{fig:M_T_a}, it appears as if U$_7$Te$_{12}$ underwent two ferromagnetic transitions along the $a$-axis and in the $c$-axis, respectively.
For the magnetic susceptibility defined as $M/H$ along the $a$-axis, a ferromagnetic transition is observed at $T_{\mathrm{C}}$ = 48 K with a steep increase in magnetization.
Below $T_{\mathrm{C}}$, a local maximum is observed in susceptibility at $T^{\star}\approx$ 26 K.
A large increase on $c$-axis magnetic susceptibility is seen below $T_{\mathrm{C}}$ and an inflection point is seen at $T^{\star}$.
This increase is observed even in higher magnetic fields indicating a ferromagnetic-like origin.
By increasing $H$ along the $a$-axis, the local maximum around $T^{\star}$ in magnetic susceptibility is suppressed (inset of Fig. \ref{fig:susc}).
The polarization of ferromagnetically ordered magnetic moments in high magnetic fields can result in an additional contribution to magnetic susceptibility which may appear as an additional shift of the local maximum at $T^{\star}$.

U$_7$Te$_{12}$ is a soft ferromagnet with a small magnetic coercivity for a magnetic field applied along the $a$-axis.
Magnetization curves for a magnetic field applied  along the $a$-axis (Fig. \ref{fig:M_H}) show almost no hysteresis down to 4.2 K,
and quickly become nearly saturated, with a magnetization of 1.41 $\mu_{\mathrm{B}}$/U at 7 T. 
Spontaneous magnetization is observed below $T_{\mathrm{C}}$, but the magnetization curve at 50 K shows an `S'-shape characteristic just above $T_{\rm C}$.
Although the $M$-$H$ data near $T_{\rm C}$ were analyzed by the the generalized Arrott--Noakes relationship $\left(H/M\right)^{1/\gamma}=(T-T_{\rm C})/T_1+(M/M_1)^{1/\beta}$ with free parameters of $T_1$, and $M_1$. The fitting could not yield reasonable critical exponents in the temperature region.
It may be due to the geometrically frustrated magnetism or competing interactions between in-plane and out-of-plane magnetism.

In the case of $H\parallel c$, the magnetization curve (Fig. \ref{fig:M_H}) exhibits a rectangular shape below $T^{\star}$.
The saturation magnetization at 4.2 K is observed to be 0.89 $\mu_{\mathrm{B}}$/U at 7 T, which is nearly half the value for the $a$-axis.
Similarly, the magnetization curve at 30 K above $T^{\star}$ shows an `S'-like shape again (Fig. \ref{fig:M_H}(b)), except for the low-field region near zero field.
In the field region above $T^{\star}$ but below $T_{\rm C}$, the $M$-$H$ curve exhibits a tiny remanent magnetization and a small coercive field (inset of the left figure in Fig. \ref{fig:M_H}), which completely disappears above $T_{\rm C}$.
The small remanent magnetization suggests a minor contribution from the $a$-axis due to misalignment.

Paramagnetic susceptibility in the high-temperature region above 150 K for the $a$-axis obeys the Curie-Weiss law, as shown in Fig. \ref{fig:susc}.
The Curie-Weiss fitting yields an effective moment $\mu_{\rm eff}$ of 1.23 $\mu_{\mathrm{B}}$/U assuming a uniform moment and Weiss temperature $\Theta_{a}$ of 63 K.
In the case of $H\parallel c$, since the inverse magnetic susceptibility is not linear, the modified Curie-Weiss law is required, which yields $\chi_{0} = 3.7\times10^{-8}$ m$^{3}$/mol, $\mu_{\mathrm{eff}}$ = 0.92 $\mu_{\mathrm{B}}$/U, and $\Theta_{c}$ = 36 K.
Magnetic susceptibility for the $c$-axis follows the modified Curie-Weiss law down to $\approx$ 60 K.
For both directions, the effective magnetic moment is much smaller than that for free U$^{3+}$/U$^{4+}$ ion.
Positive $\Theta$ means that ferromagnetic interactions are dominant in the system, and $\Theta_{a} >\Theta_{c}$ suggests the in-plane interactions are stronger.

Suski \citep{Suski1972} previously discussed the possibility of different magnetic moments by assuming $^{3}$H$_{4}$ term for U$^{4+}$ ($5f^2$) for the uranium but different splitting based on the two types of polyhedral U sites, shown in Fig.~\ref{fig:structure}.
However, because U$_{7}$Te$_{12}$ can be formally considered as U$^{4+}_3$U$^{3+}_4$Te$_{12}^{2-}$ from the charge neutral principle, valence conditions are probably needed, e.g. U$^{3+}$ state on the U1 site, and U$^{3+}$ or U$^{4+}$ state assigned on the U2 and U3 sites, respectively.
Additional experiments are desirable to determine the CEF and valence states.
The magnetic anisotropy is evidently expected to be site-dependent in such a situation, but based on the ratio of magnetic moments $M_{a}/M_{c}$(2 K, 7 T) $\approx$ 4/3 and multiplicity of uranium sites we expect uraniums on U1 and U2 or U3 to be ordering in-plane at $T_{\mathrm{C}}$ while uraniums on U2 or U3 site are responsible for the $c$-axis magnetization increase at $T^{\star}$.

\subsection{Heat capacity}
\begin{figure}[!thb]
\includegraphics[width=8.5cm]{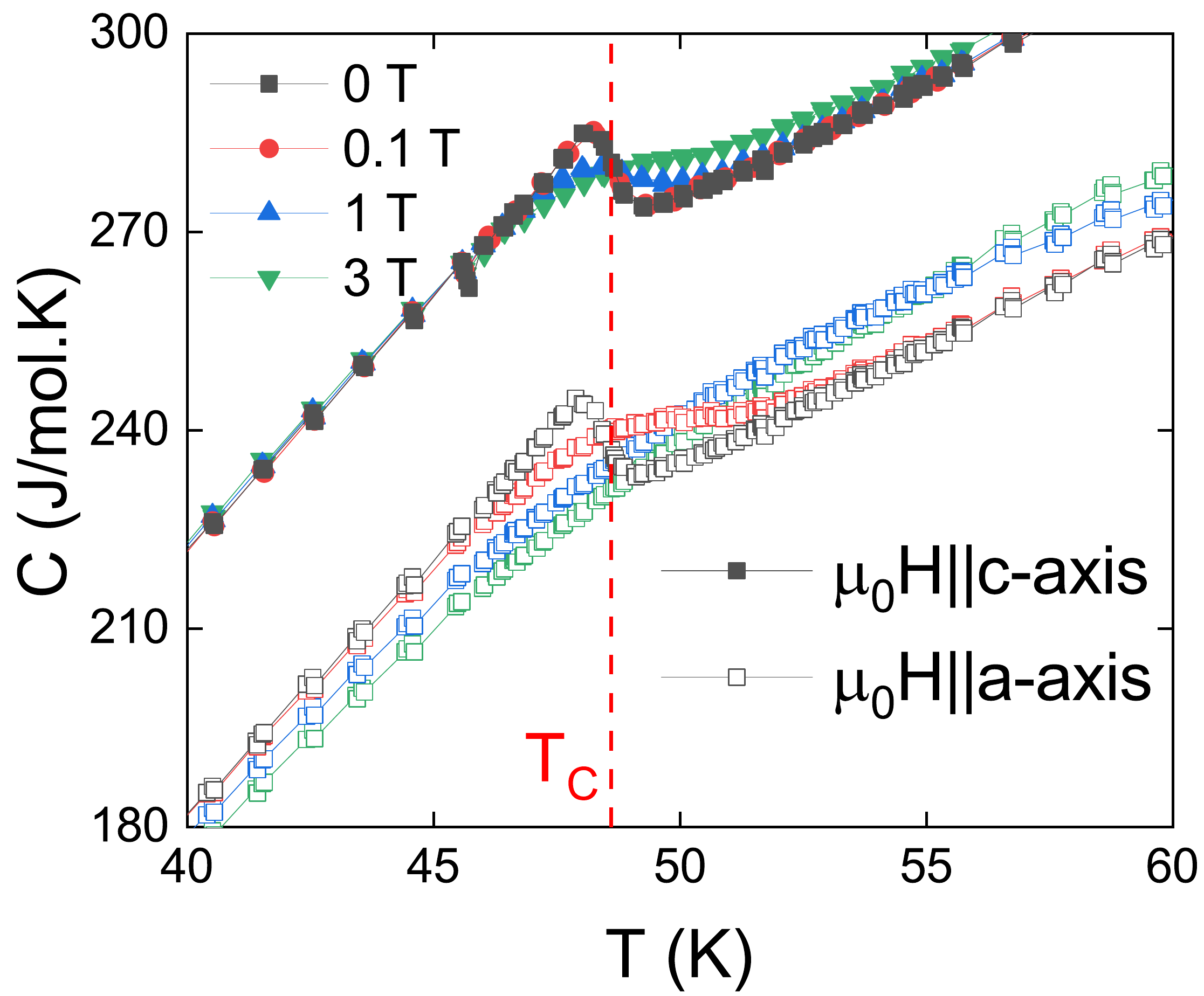}
\caption{\label{fig:HC_1}
Heat capacity of U$_{7}$Te$_{12}$ in different magnetic applied along the $c$-axis and $a$-axis.
Data for $H\parallel a$ (empty symbols) are shifted downward by 40 J/mol$\cdot$K for clarity.}
\end{figure}
\begin{figure} [!thb]
\includegraphics[width=8.5cm]{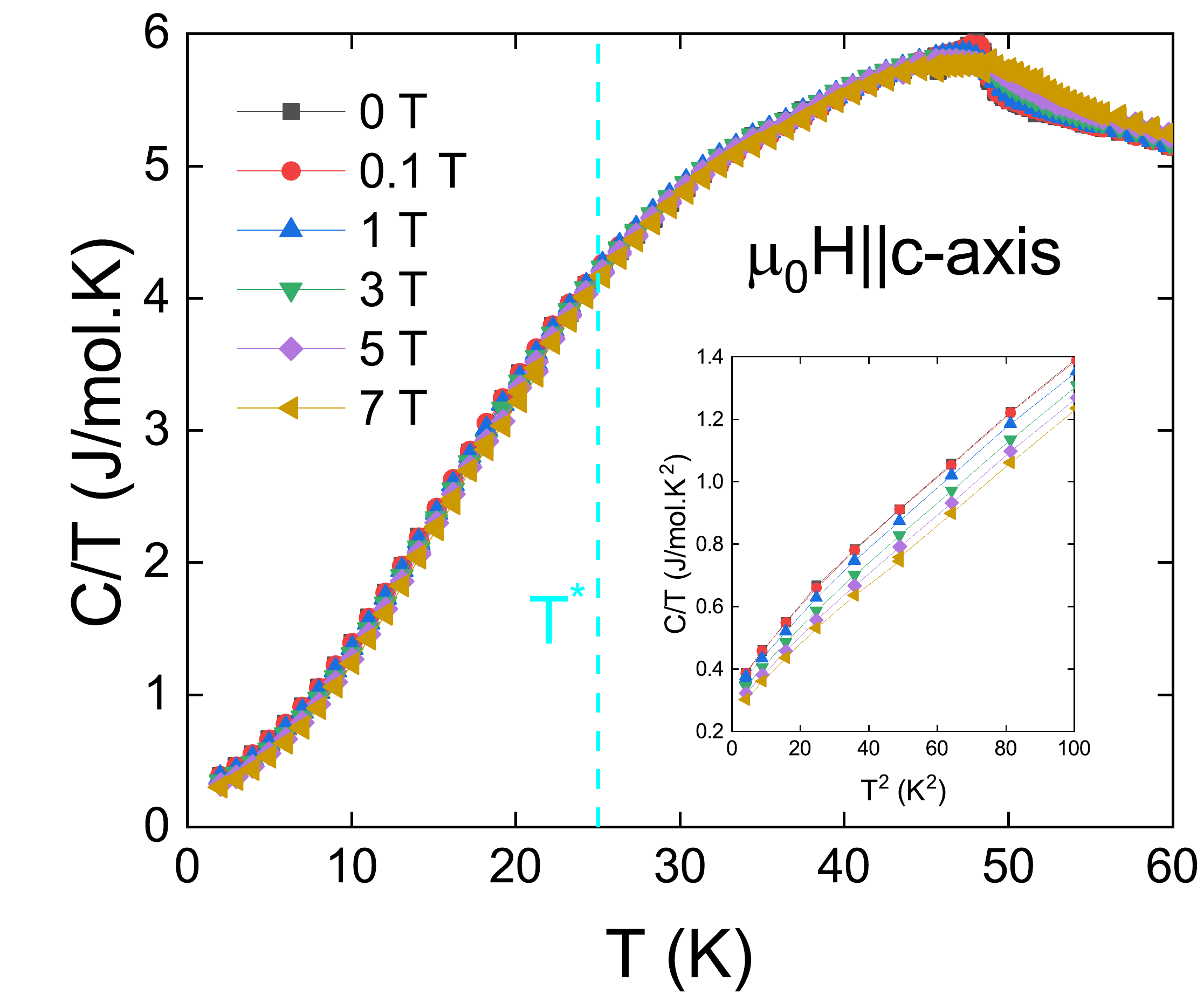}
\caption{\label{fig:HC_2_c}
Plot of $C/T$ vs. $T$ of U$_{7}$Te$_{12}$ in the magnetic field applied along the $c$-axis.
The inset shows the low-temperature part of $C/T$ vs. $T^{2}$ to determine the Sommerfeld coefficient.
}
\end{figure}
 
Specific heat measurements were performed to examine the magnetic anomalies at $T_{\rm C}$ and $T^{\star}$.
The results for various magnetic fields applied in both directions are shown in Figs. \ref{fig:HC_1} and  \ref{fig:HC_2_c}.
Only one phase transition is observed at 48.6 K corresponding to $T_{\mathrm C}$.
Concurrently, no $\lambda$-type anomaly is observed around $T^{\star}$, indicating that the anomaly at $T^{\star}$ is a magnetic crossover.

As shown in Fig. \ref{fig:HC_1}, in the case of $H\parallel a$, the $\lambda$-type anomaly at $T_{\rm C}$ is easily blurred by a tiny external field of 0.1 T.
Meanwhile, in the case of $H\parallel c$, the anomaly remains at 0.1 T, despite being blurred by an $a$-projected field due to a field misalignment within several degrees.
The Sommerfeld coefficient well below $T_{\rm C}$ is extrapolated to be $\gamma = 48$ mJ/mol$\cdot \mathrm{K}^{2}$ from data below 10 K, which is similar to the $\gamma$ values well below $T_{\rm C}$ in other  ferromagnetic uranium compounds with itinerant $5f$ electrons (e.g., UCoAl $\gamma = 48$ mJ/mol$\cdot \mathrm{K}^{2}$ \citep{Matsuda}, UCu$_{2}$Ge$_{2}$ $\gamma = 30$ mJ/mol$\cdot \mathrm{K}^{2}$ \citep{Matsuda2007}, UGe$_{2}$ $\gamma = 35$ mJ/mol$\cdot \mathrm{K}^{2}$ \citep{Onuki1992}, UIr $\gamma = 50$ mJ/mol$\cdot \mathrm{K}^{2}$ \citep{Galatanu2004}, U$_{5}$Sb$_{4}$ $\gamma = 37$ mJ/mol$\cdot \mathrm{K}^{2}$ \citep{Paixao1994}).

In the $C/T$ vs. $T$ plot as shown in Fig. \ref{fig:HC_2_c}, instead of a $\lambda$-type anomaly, there is a sign of hump anomaly that spreads widely around $T^{\star}=26$ K, which does not seem to be affected by external fields along the $c$ axis and no change was observed for field applied along the $a$-axis.
For a crossover from in-plane ferromagnetism to $c$-axis ferromagnetism of uniform moments, applying an external magnetic field will change $C/T$ vs. $T$ at the crossover, but it is not the case in U$_{7}$Te$_{12}$.
Therefore, the crossover at $T^{\star}$ is most likely a crossover from a paramagnetic state to a ferromagnetic state with an out-of-plane component,  in the background of in-plane ferromagnetism.
Namely, it suggests the presence of partially disordered (paramagnetic) uranium sites in the intermediate region of $T^{\star}<T<T_{\rm C}$. 
It should be also noted that the $T^{\star}$ is reduced by the fields applied along the $a$-axis in electrical resistivity and magnetization.
As shown in the inset of Fig. \ref{fig:HC_2_c}, the $\gamma$ value is reduced by an external field along the $c$ axis, but it is almost independent of $H$ along the $a$-axis.
This indicates that the low-temperature state is only sensitive to the magnetic field in the $c$ direction.

\subsection{Electrical resistivity}
\begin{figure}[bht]
\includegraphics[scale=0.7]{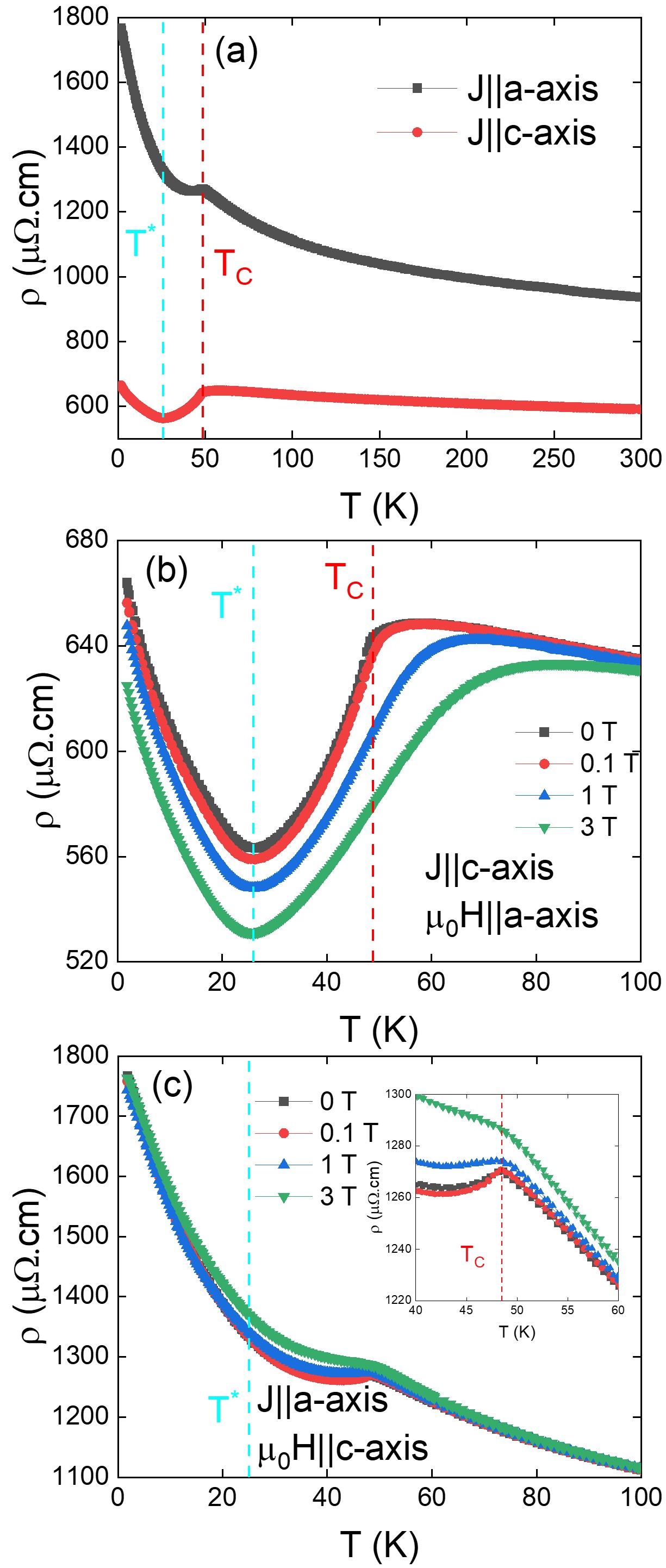}
\caption{\label{fig:R_T}
(a) Temperature dependence of electrical resistivity $\rho_c$ and $\rho_a$ for U$_{7}$Te$_{12}$ measured with the electrical current ($I$) applied along the $c$- and $a$-axes, respectively.
(b) $\rho_c(T)$ with applying fields along the $a$-axis.
(c) $\rho_a(T)$ with applying fields along the $c$-axis.
The inset shows a magnified $\rho$-$T$ plot around $T_{\mathrm C}$.
$T_{\mathrm C}$ and $T^{\star}$ are highlighted by red and blue lines, respectively.
}
\end{figure}

As shown in Fig. \ref{fig:R_T}(a), U$_{7}$Te$_{12}$ exhibits a semimetallic behavior.
The electrical resistivity $\rho(T)$ is much lower than the reported value of $\rho({\rm 300\ K})\simeq$3 m$\Omega\cdot$cm \cite{Tougait1998}, indicating the high-quality of our crystal.
The $\rho_c(T)$ in the $c$-axis direction is smaller than $\rho_a(T)$ in the $a$-direction, confirming the high anisotropy of U$_{7}$Te$_{12}$.
Both $\rho_a(T)$ and $\rho_c(T)$ show a gradual increase with decreasing temperature from 300 K.
At $T_{\rm C}=48$ K, $\rho_a(T)$ shows a local maximum and $\rho_c(T)$ shows a kink anomaly.
Notably, around $\sim$60 K above $T_{\rm C}$, a broad maximum is seen. Below 48 K, $\rho_a(T)$ keeps increasing without any other anomalies.
For $\rho_c(T)$, a sharp decrease is observed below $T_{\rm C}$, but increases sharply below $T^{\star}=26$ K.
The $\rho_c(T)$ below $T^{\star}$ does not follow the Arrhenius-type $T$-dependence, i.e., it is not a semiconducting behavior with a fixed energy gap.
It may be a half-gapped semimetallic state induced by the ferromagnetic crossover.

Figures \ref{fig:R_T}(b) and \ref{fig:R_T}(c) show the magnetic field dependence of $\rho_c(T)$ and $\rho_a(T)$ measured with applying magnetic fields along the perpendicular directions to the current ($I$) directions.
For $\rho_c$, as shown in Fig. \ref{fig:R_T}(b), when the magnetic field is applied along the $a$-axis, the ferromagnetic transition at $T_{\rm C}(0)=48$ K moves to higher temperatures, and it is easily smeared out as magnetic fluctuations are suppressed.
In other words, the negative magnetoresistance effect of $\rho_{c}$ above $T_{\rm C}$ is significant in the in-plane field direction.
The $T^{\star}$ slightly moves to lower temperatures with an increasing magnetic field, which is similar to the behavior observed in magnetic susceptibility in $H\parallel a$.
The $\rho_c(T)$ below $T^{\star}$ does not show any significant magnetoresistance effect by applying fields along the $a$ axis.
Alternatively, in $H\parallel c$ shown in Fig. \ref{fig:R_T}(c), the $T_{\rm C}$ does not shift significantly below $\sim$3 T.
There is no significant magnetoresistance effect in this field direction.
The sample was easily moved from the fixed position in the higher fields along the hard magnetic axis due to the strong magnetic torque.
Therefore, the data above 3 T were unable to be collected under these conditions.

\section{\label{sec:level3}Discussions}

To begin with the discussion, the main experimental results are summarized as follows.
Two anomalies at $T_{\rm C}=48$ K and around $T^{\star}=26$ K are observed in U$_7$Te$_{12}$.
A ferromagnetic ordering occurs at $T_{\rm C}$ where the ferromagnetic component emerges in the basal plane.
In the ferromagnetic state below $T_{\rm C}$, an additional ferromagnetic crossover along the $c$-axis occurs around $T^{\star}$.
In the specific heat measurement, a clear $\lambda$-type anomaly is observed at $T_{\rm C}$, whereas a broad hump is barely visible around $T^{\star}$.
Such an anisotropic ferromagnetic response along the $a$- and $c$-axes can be due to the multiple uranium sites of U1, U2, and U3.
As shown in Sec. \ref{sec:level2}B, the in-plane ferromagnetism below $T_{\rm C}$ can be ascribed to the four out of the seven uranium moments, while the rest uranium moments would be responsible for the ferromagnetic crossover along the $c$ axis below $T^{\star}$.
Based on these observations, the nature and origin of the anomaly at $T^{\star}$ would be discussed.

In general, a crossover between paramagnetic and ferromagnetic states is often induced by external magnetic fields.
This occurs in U$_7$Te$_{12}$ even at zero magnetic field.
The ferromagnetic transition at $T_{\rm C}$ is a second-order transition, breaking the time-reversal symmetry.

Considering the in-plane ferromagnetic ordering at the $T_{\mathrm{C}}$, the symmetry of U1, U2, and U3 positions denoted in Table \ref{tab:table2} seems to be reduced to $m'..$ by time-reversal symmetry breaking.  
If this were the case, the appearance of ferromagnetic moments along the $c$-axis below $T^{\star}$ broke the $m'$ symmetry to the lowest symmetry 1, which should induce a phase transition at $T^{\star}$.
This contradicts the observed crossover behavior below $T^{\star}$, indicating that the symmetry at the uranium sites is already lowered to 1 at $T_{\mathrm{C}}$.
At the present, the origin of this symmetry lowering to 1 at $T_{\mathrm{C}}$ is unknown.

Let us consider the in-plane ferromagnetic structure below $T_{\rm C}$ again.
Considering the in-plane arrangement of uranium atoms, the U1 moments can form a collinear ferromagnetic structure without the cost of anisotropy energy.
However for U2 sites, because the local easy axis is not parallel between the neighboring sites, a collinear ferromagnetic structure costs the anisotropy energy.
Thus, non-colinear ferromagnetic states, such as 2-in-1-out (1-in-2-out), are considered more favorable.

The same situation might be realized for the U3 site.
In reality, the interaction between the out-of-plane site and the uranium moments of the other sites must be considered further.
As a result, the overall ferromagnetic structure seems to be more complex.
To identify the magnetic structure, it is necessary to perform neutron scattering experiments.

\section{Conclusion}
We have succeeded in growing the single crystals of U$_{7}$Te$_{12}$ and characterized the physical properties.
The ferromagnetic transition is observed at $T_{\rm C}=48$ K in different measurements with magnetic moment parallel to the basal plane.
The crossover is observed around $T^{\star}=26$ K and is characterized by an increase of magnetic moment along the $c$-axis and no dependence on the magnetic field.
We explain the low-temperature state below $T^{\star}$ as the ferromagnetic state with magnetic moment along the $c$-axis or canted from the $c$-axis.
 We propose that ordered magnetic moments originate from uranium at different crystallographic positions giving rise to ordering at $T_{\rm C}$ and $T^{\star}$.

\section{Acknowledgement}
We wish to thank C. Tabata, K. Kaneko, and M.-T. Suzuki, for the helpful discussion.
A part of this work was supported by JSPS KAKENHI Grant Nos. JP20KK0061, and JP20K20905.
This work was also partly supported by the JAEA Fund for Exploratory Researches (Houga fund) and REIMEI Research Program.

\bibliography{liter}

\end{document}